\documentclass[12pt,a4paper]{article}
\usepackage{cite}
\usepackage{graphicx}
\usepackage{longtable}
\usepackage{booktabs}
\usepackage{amsmath}
\usepackage{amssymb}
\usepackage{bm}
\usepackage{breqn}
\usepackage{color}
\usepackage{float}
\usepackage{graphicx}
\usepackage{hyperref}
\usepackage{indentfirst}
\usepackage{multicol}
\usepackage{subfigure}    
\usepackage{multirow}
\usepackage[normalem]{ulem}
\usepackage{appendix}
\usepackage{geometry}
\usepackage{authblk}
\usepackage{fancyhdr}
\title{Detecting Suspected Epidemic Cases Using Trajectory Big Data}

\date{} 

\author[1]{Chuansai Zhou}
\author[2]{Wen Yuan}
\author[2]{Jun Wang}
\author[3]{Haiyong Xu}
\author[3]{Yong Jiang}
\author[4]{Xinmin Wang\mbox{${}^{1,}$}}
\author[4]{Qiuzi Han Wen \thanks{Corresponding author: qiuzi.wh@pku.edu.cn}\mbox{${}^{1,}$}}
\author[4]{Pingwen Zhang \thanks{Corresponding author: pzhang@pku.edu.cn}\mbox{${}^{1,}$}}
\affil[1]{School of Mathematical Sciences, Peking University, Beijing 100871, China}
\affil[2]{Academy for Advanced Interdisciplinary Studies, Peking University, Beijing, China}
\affil[3]{China Mobile Information Technology Co., Ltd., Beijing, China}
\affil[4]{National Engineering Laboratory for Big Data Analysis and Application, Peking University, Beijing, China}

\begin{document}             
	
	\maketitle                   
	
	\begin{abstract}
		
		Emerging infectious diseases are existential threats to human health and global stability. The recent outbreaks of the novel coronavirus COVID-19 have rapidly formed a global pandemic, causing hundreds of thousands of infections and huge economic loss. The WHO declares that more precise measures to track, detect and isolate infected people are among the most effective means to quickly contain the outbreak. Based on trajectory provided by the big data and the mean field theory, we establish an aggregated risk mean field that contains information of all risk-spreading particles by proposing a spatio-temporal model named HiRES risk map. It has dynamic fine spatial resolution and high computation efficiency enabling fast update. We then propose an objective individual epidemic risk scoring model named HiRES-p based on HiRES risk maps, and use it to develop statistical inference and machine learning methods for detecting suspected epidemic-infected individuals. We conduct numerical experiments by applying the proposed methods to study the early outbreak of COVID-19 in China. Results show that the HiRES risk map has strong ability in capturing global trend and local variability of the epidemic risk, thus can be applied to monitor epidemic risk at country, province, city and community levels, as well as at specific high-risk locations such as hospital and station. HiRES-p score seems to be  an effective measurement of personal epidemic risk. The accuracy of both detecting methods are above 90\% when the population infection rate is under 20\%, which indicates great application potential in epidemic risk prevention and control practice.

		
		
	\end{abstract}
	

	\section{Introduction}
	
	One of our greatest challenges is the continuing global impact of infectious diseases. On March 11th of 2020, The World Health Organization declared spread of the novel coronavirus COVID-19 is a global pandemic, when 118,000 confirmed cases of COVID-19 were found in 114 countries, with 4,291 deaths\cite{times}. Although SARS-CoV-2, the virus that causes COVID-19, has been found to have lower case fatality rate than either SARS or Middle East respiratory syndrome-realted coronavirus (MERS-CoV)\cite{zhong}, the sheer speed of its geographical expansion and surge in numbers of confirmed cases severely impact public health system. Human-to-human transmission seems to be the main method of transmission for SARS-CoV-2, according to CDC of the United States\cite{cdc_usa}. The human-to-human transmission routes of SARS-CoV-2 include direct transmission, such as cough, sneeze, droplet inhalation transmission, and contact transmission, such as the contact with oral, nasal, and eye mucous membranes\cite{Peng_2020}. But limitation of our knowledge on SARS-CoV-2 and diversified symptoms of the infected patients complicate the diagnosis of COVID-19\cite{zhong}\cite{co-infection}. In addition, a significant percentage of infected patients, ranging from 18\% to 50\% as estimated in different research, are asymptomatic or with mild symptoms, but they could still be highly contagious \cite{qiu2020covert}. Therefore, early identification of infected individuals and blocking their transmission paths are keys to effectively stop virus from spreading. 
	
	Classic epidemic models such as SIR or exponential growth model use mathematical and statistical approaches to quantify the dynamic mechanism of epidemic transmission and to predict the size of the infected population. In these epidemic models, basic reproduction number $R_0$ is the key parameter for quantifying the virus epidemic. Liu et al. reviewed published studies on estimation of $R_0$ of COVID-19 extracted from PubMed, bioRxiv and Google Scholar during January 1st to February 7th, 2020\cite{Liut_2020}. Results show that the estimated $R_0$ varies from 2.2 to 6.49, and estimations obtained from the mechanism-based dynamic modeling is significantly higher than those obtained using statistical methods based on exponential growth model. Lack of robustness in estimation of the basic reproduction number may result in misleading estimates of the epidemic trend of COVID-19 as well as the size of infected population. In addition, these epidemic models cannot predict the infection status of a specific individual, thus can only play a limited role in practice of epidemic prevention and control. Epidemiology investigation used to be the main method of exploring transmission process of infected individuals, but ``omissions and errors in previous activities can occur when the investigation is performed through only a proxy interview with the patient"\cite{Korea_cdc}.
	
	With the development of emerging technologies such as cloud computing, big data and artificial intelligence, epidemiological research as well as epidemic prevention and control methods are undergoing innovation. For example, based on search engine query data, Google developed an approach, i.e., the Google Flu Trends (GFT) model, for detecting influenza epidemics through monitoring health-seeking behavior in the form of queries to online search engines. It provides estimates on the degree of influenza activities at weekly and regional scale for the United States, with a reporting lag about one day\cite{gflu_nature}\cite{gft_2011}. Another example is more recent. South Korea's response to the outbreak of COVID-19 has been applauded by a number of international health experts \cite{korea_news}. Its outstanding achievements in the prevention and control of the COVID-19 epidemic largely depend on the comprehensive application of multi-source data, such as data from medical facility records, Global Positioning System, card transactions, and closed-circuit television. Methods that can objectively verify the patient’s claims are adopted for conducting COVID-19 contact investigations in South Korea, which have collected more accurate and timely information on the location, time of exposure, and details of the transmission situations\cite{Korea_cdc}.
	
	In this research, based on trajectory big data and mean field theory, we firstly propose a high-resolution spatio-temporal model, namely HiRES, for the risk assessment of epidemic disease with human-to-human transmission. HiRES has a fine spatial resolution and high computation efficiency that can support real-time monitoring in practice. Then using the epidemic risk maps produced by HiRES model, we propose a personal infection risk scoring model, namely HiRES-p, to obtain objectively quantified risk of infection for every authorized individual. Based on these risk scores, we develop methods using statistical inference and machine learning approach respectively to detect early infection of suspected cases. We apply our methods to investigate early outbreak of COVID-19 in China during Jan 1st, 2020 to Jan 28th, 2020.

	\section {Trajectory big data}
	
	Trajectory data refers to a sequence of geographic locations with timestamps. It is a series of sampling points from the trajectory of continuously moving objects\cite{feng2016survey}. According to the different acquisition methods, trajectory data can be divided into the following five categories: GPS (global positioning system) based, wireless network signaling based, geo-social network based, RFID (radio frequency identification) based and Wi-fi based\cite{mazimpaka2016trajectory}. On the other hand, according to the type of moving objects, it can be classified into human movement trajectory, vehicle trajectory, animal trajectory and natural phenomenon movement trajectory\cite{zheng2015trajectory}.
	
	Distinctive features of trajectory data include large-volume, real-time and diversity. In addition, trajectory data in general carries a lot of information for data mining and scientific analysis, thus is a very important type of big data. The spatio-temporal variability is the basic feature of trajectory data since it contains information of spatial and temporal dimensions. Due to the variety of trajectory data sources, the trajectory data is sampled at different frequencies. In addition, sampling of continuously moving objects may introduce more biases and difficulties in data processing. Considering these characteristics, analysis and mining of trajectory big data must balance between discovering the spatio-temporal structure and variability and avoiding introducing excessive noise. 
	
	Common methods for analyzing the trajectory big data include pattern mining, clustering, machine learning, spatio-temporal modeling, etc. For example, \cite{xia2018exploring}\cite{gong2016inferring}\cite{sakuma2019efficient} discover common characteristic of objects' movement behaviors based on trajectory data by pattern mining;  \cite{zheng2010understanding}\cite{bashir2007object} use traditional machine learning based on feature engineering, while \cite{zhang2019classifying}\cite{liu2019spatio} apply neural network-based models, to classify trajectory data with different characteristics; \cite{olive2019trajectory}\cite{jung2019outlier}\cite{pelekis2017temporal} apply clustering methods to calculate data similarity in order to find abnormal trajectories and to mine trajectory patterns; recent works in \cite{kong2016urban}\cite{wang2019crash} predict related variables in both spatial and temporal dimensions based on existing data. From the perspective of application scenario, trajectory big data has been applied to traffic prediction \cite{kong2016urban}\cite{wang2019crash}, route recommendation, travel services\cite{kong2017time}, travel behavior analysis, smart city management\cite{xia2018exploring} and so on.
	
	Trajectory big data used in this study is calculated and analyzed within the Phoenix Tree Platform which is a centralized big data PaaS platform operated by one of the global Tele Communication Company China Mobile. To ensure data security and protect personal information, we sanitize the datasets so that the direct link between the trajectory and the specific subscriber cannot be identified. In the training datasets, subscribers are labeled as `confirmed' or `normal'. The use and disclosure of data have strictly followed data and information management regulations in China and China Mobile's data management requirements and obtain its approval and individual authorization\cite{creemers2016cybersecurity}\cite{chapter2006law}\cite{national2007emergency}\cite{national2011emergency}.
	
	\section{Mathematical modeling of epidemic risk}
	
	We develop mathematical models to quantify regional epidemic risk and personal infection risk. Firstly, based on the mean field theory, we propose a spatio-temporal model, namely HiRES, to provide high-resolution estimation of regional epidemic risk based on trajectory data of confirmed infections and obtain HiRES epidemic risk maps through mapping the estimated spatial epidemic risk to the corresponding base station cells. The main idea of the mean field theory is to focus on one particle and it assumes that the most important contribution to the interactions of such particle with its neighboring particles is determined by the mean field due to the neighboring particles\cite{mft-2020}. Based on this theory, instead of investigating how each confirmed case, i.e., a risk-spreading particle, interact with the rest of population, which is impossible or computationally costly, we establish an aggregated risk mean field that contains information of all risk spreading particles. Then based on this epidemic risk mean field, we propose a scoring model HiRES-p to estimate the personal epidemic risk of each individual using the HiRES maps and his/her trajectory data. Finally, based on the HiRES-p scores, we develop detection methods using both statistical inference approach and machine learning approach to detect and classify suspected individuals.

	
	\subsection{Notations}
	
	We start by introducing key notations used in this paper.
	
	${t_0,t_1,\cdots,t_k}$: time period; 
	
	$\hat{t}_k=t_k-t_{k-1}$: basic temporal resolution;
	
	$\hat{x}_{l,m}$: identification of a communication base station cell (abbreviated as BSC hereafter), where $l$ means area number, m is symbol number of station in this area. Every spatial coverage of BSC varies from 500m to 1km approximately. We will use $\hat{x}_{l,m}$ indiscriminately to represent station and treat as one pixel in our epidemic risk map; 
	
	$F(\hat{x}_{l,m}, \hat{t}_k)$: base function at station $\hat{x}_{l,m}$ during time period $\hat{t}_k$;
	
	$\bar{F} (\hat{x}_{l,m}, \hat{t}_k)$: estimated epidemic risk at BSC $\hat{x}_{l,m}$ during time period $\hat{t}_k$;
	
	$f_p(\hat{x}_{l,m}, \hat{t}_k)$: subscriber $p$’s proportion of stay-time at BSC $\hat{x}_{l,m}$ during $\hat{t}_k$; 
	
	$y_p(\hat{t}_k)$:  subscriber $p$’s personal base risk;
	
	$\tilde{y}_p(t_k)$: subscriber $p$’s personal risk score at time $t_k$;
	
	$\delta_s$: incubation decay coefficient which represents degree of infection impact of a confirmed case on a BSC;
	
	$\Gamma_i$: outdoor decay coefficient of virus;
	
	$\gamma_i$: decay coefficient of virus.

	\subsection{HiRES model and epidemic risk map}
	
	\label{3.2}
	
	The spreading risk of an infectious disease with human-to-human as the main route of transmission comes from human movements and interaction. Thus high-resolution trajectory data of infected patients can well represent the spreading spatio-temporal risk of the virus. Put it in the context of mathematics, given the trajectory data of an infectious patient $p$, i.e., confirmed case,  we define his/her proportion of stay at certain base station cell $\hat{x}_{l,m}$ during a given time period $\hat{t}_k$ by $f_p(\hat{x}_{l,m},\hat{t}_k)$, which satisfies $\sum_{l,m} f_p(\hat{x}_{l,m},\hat{t}_k) = 1$. We define our base function $F$ as follows:
	
	\[F( \hat{x}_{l,m}, \hat{t}_k) = \sum_{p=1}^P \delta_{s(p,\hat{t}_k)} f_p(\hat{x}_{l,m}, \hat{t}_k)
	\]
	where $P$ is the total number of confirmed cases, $\delta_{s(p,\hat{t}_k)}$ is incubation decay coefficient as defined in Sect. 3.1. Since each confirmed case has different confirmed date, let $ s(p,\hat{t}_k)$ be a function of subscriber and time interval, which means the number of unit of time from $t_k$ to confirmed date. For example, if we choose time interval as one day,  subscriber $p$ diagnosed on 15th Jan, we intend to calculate basic value at 13th Jan ($t_k = 13th, t_{k-1} = 12th$), then $ s(p,\hat{t}_k)$ = 2.

	We consider the median length of stay in hospital as a proxy for time to cure, and we assume that the patient would become temporarily immune to the virus when healed. Thus after recovery, confirmed cases will be removed from the confirmed group and their stay at BSCs will not be counted into the risk map anymore.
	
	Based on the trajectories of all confirmed cases during a certain period, for every region covered by a BSC, its epidemic risk can be calculated by:
	$$
	\bar{F}(\hat{x}_{l,m})=\sum_{k}\Gamma_k F(\hat{x}_{l,m},\hat{t}_k)
	$$
	where $k$ depends on the survival time of virus. $\Gamma_k$ is the decay coefficients. Thus, we obtained our epidemic risk map, namely, HiRES risk map, $\{\bar{F}(\hat{x}_{l,m},lat_{l,m}, lon_{l,m})\}$, where $(lat_{l,m}, lon_{l,m})$ refers to the geographical information (latitude and longitude) of the BSC. The spatial coverage and resolution of this HiRES map is determined by the spatial distribution and coverage/radius of all the BSCs covered by the trajectory data.  
	
	\subsection{HiRES-p model and detection of suspected cases}
	
	\label{3.3}
	
	In the practice of epidemic prevention and control, one of the most important and difficult problems is to detect suspected cases, i.e., people are highly likely infected or to be infected in the near future. Based on HiRES risk maps, we develop a personal epidemic risk scoring model HiRES-p, which can explicitly quantified risk of infection for every authorized individual. The infection risk of an individual during a certain time interval is defined by his/her accumulated trajectory risk given by the HiRES risk maps. During a time interval $\hat{t}_k $, the HiRES-p personal base risk score is formulated as follows:
	
	$$
	y_p(\hat{t}_k)=\sum_{\hat{x}_{l,m}}\sum_{\hat{t}_k}\gamma_k f(\hat{x}_{l,m},\hat{t}_k)F(\hat{x}_{l,m},\hat{t}_k)
	$$
	where $k$ depends on the survival time of virus. Most infectious viruses have an incubation period, when we evaluate the infection risk of an individual at certain time point, it's not enough to only consider risk arising from the most recent exposure. Historical personal risk score exposure during the incubation period also plays a crucial role. Thus we define the HiRES-p personal risk score at time $t_k$ by the maximum of his/her epidemic risk score series obtained during the past incubation period. That is, the HiRES-p scoring model can be given by:
	$$
	\tilde{y}_p(t_k) = max(y_p(\hat{t}_k), y_p(\hat{t}_{k-1}),\cdots, y_p(\hat{t}_{k-T})) 
	$$
	where $T$ is the incubation period of the epidemic virus. By definition, the distribution of the HiRES-p score for an individual will follow a generalized extreme value distribution (GEV). Generally speaking, HiRES-p risk score is defined by a function of historical personal risk exposures during the virus incubation period, thus can take other form besides taking maximum. In machine learning based method introduced later in this article, the analytic form of this function defines data feature. 
	
	To detect suspected cases, we firstly use statistical inference approach to formulate the detection problem into a hypothesis testing problem. Assume the HiRES-p score of a normal individual follows a GEV distribution with its cumulative distribution function given by $F_0(x;\mu,\sigma,\eta)$. The real-time HiRES-p score $ \tilde{y}_p(t_k)$ is considered as a sample from certain GEV distribution. Numerical experiments indicate that the probability distribution of HiRES-p for the normal group is significantly different from that of the infected group. Thus, detection of suspected cases is equivalent to test whether sample $ \tilde{y}_p(t_k)$ comes from the distribution under the null hypothesis, i.e., $F_0$. We calculate the P-value of this hypothesis testing problem based on the null distribution $F_0$, and reject the null hypothesis if it is small at certain significance level, say 5\%, which indicates a successful detection of suspected case. The theoretical detection rate (DR) is defined by the power of the test and false alarm rate (FAR) is relevant to the significance level chosen for the hypothesis test. let $\alpha$ denote type I error and $\beta$ denote type II error, we have the following theoretical results on DR and FAR:
	
	$$
	\left\{
	\begin{array}{lcl}
	DR=1-\beta \\
	FAR=\alpha
	\end{array}
	\right.
	$$  
	
	From the big-data perspective, detection of suspected cases can be formulated as a classification problem using machine learning approach. Given large group of people, we would like to divide them into two categories: suspected case and healthy case. To apply machine learning approach, we firstly need to construct the training dataset and the testing dataset with sufficient amount of trajectory records. Then we extract data features by calculating the HiRES-p score defined by certain analytic form. Given a labeled training dataset, we can train our model using common machine learning algorithms such as Decision Tree, Random Forest and Support Vector Machine. Performance of the ML methods can be evaluated using the testing dataset. Model selection should consider both accuracy and robustness.
	
	The flowchart of models and methods introduced in this section is summarized in the following figure.
	
	\begin{figure}[htbp]
		\centering
		\includegraphics[scale=0.8]{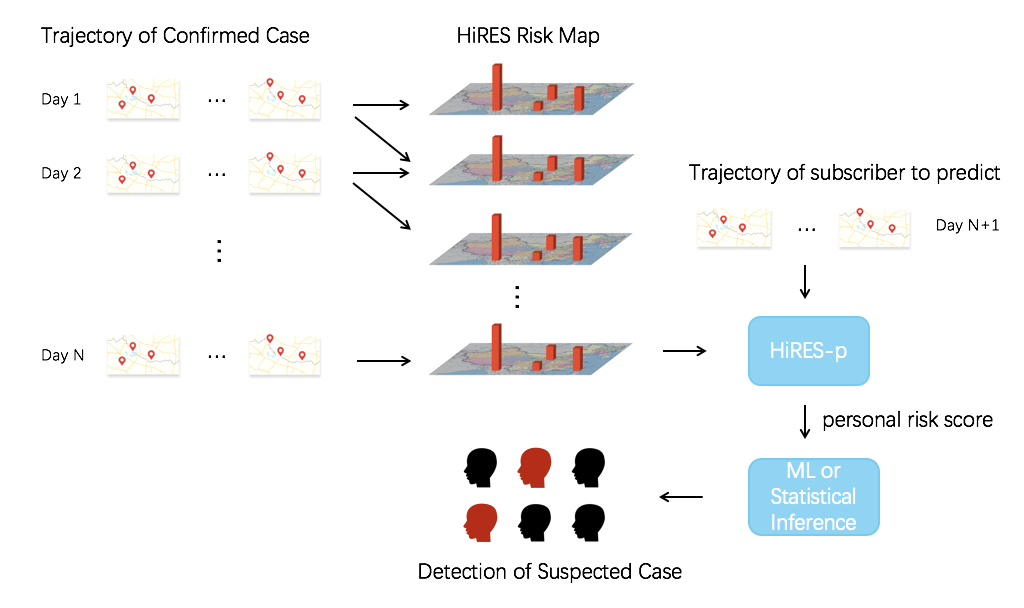}
		\caption{Architecture of HiRES and HiRES-p}
	\end{figure}
	
	\subsection{Discussions on hyperparamters}
	
	We discuss hyperparameters that are important in our method.  
	
	1. Basic temporal resolution
	
	The basic temporal resolution is chosen as one day in this research, but can be changed to fit data and research need. For convenience of expression, we will use $k$ replace $\hat{t}_k$ which means day $k$. In practice, higher resolution is preferable since faster update of personal risk enables earlier detection of suspected cases, meanwhile higher resolution demands more data and computing resources.  
	
	2. Incubation decay coefficient
	
	Incubation decay coefficient is a key parameter in calculating the HiRES risk map, thus validation of this parameter is crucial for the accuracy of the risk map. It consists of a virus decay model and an estimate of incubation period. In practice, theoretical results may not be available especially at the early outbreak of the disease. Advices from professionals are needed. In our application study, we work on early outbreak of COVID-19 in China. \cite{ref9} Linton et. al. found that the incubation period of COVID-19 falls within the range of 2-14 days with $95\%$ confidence. Based on this research, we set incubation period T=14 days in our experiments. We assume virus exponential decay follows this formula $\delta_s = e^{-\frac{1}{T+1-s}}, s=1,2,...,14$, where s represents number of days from current time to diagnosed date, and $\delta_s = 0$ when s>14.
	
	3. Decay coefficient of virus
	
	Outdoor decay coefficient measures how long virus stay active after viral shedding. Recent research find that COVID-19 was most stable on plastic and stainless steel and viable virus could be detected up to 72 hours \cite{ref1}. Thus, in our application study, epidemic risk on BSCs and personal risk score are the weighted sum of 3 daily values.
	
	\[\bar{F}(\hat{x}_{l,m}, k) = \sum_{i=0}^2 \Gamma_i F(\hat{x}_{l,m}, k-i ) \]
	where $\Gamma_i$ are decay weights outdoors. Under the assumption of viable virus being reduced by half each day, we let $\Gamma_i = 50\times 2^{2-i}$.
	
	 Decay coefficient of virus $\gamma_i$ measures rate of decreasing of infectiousness of viruses to humans and is used to calculate the HiRES-p score: 
	\[ y_p(k) = \sum_{i=0}^2 \gamma_i \sum_{\hat{x}_{l,m}} f_p(\hat{x}_{l,m}, k-i) \bar{F}(\hat{x}_{l,m}, k-i)\]
	 \cite{ref1} Neeltje et al. showed COVID-19 linearly decay under $log_{10} TCID_{50}/mL$ titer, so we set $\gamma_i = 10^{-i}$ in our application study.
	
	4. Recovery time of an infected individual
	
	Recovery time is an important index we use to decide whether an infected individual still contributes to the risk map or not. In practice, exact recovery time of each individual is hard to get. Instead, record of stay in hospital for confirmed cases can be obtained and used to estimate time to cure. Recent research find that the median length of stay in hospital is 10 days in China. The largest median length of stay in hospital is 20 days in Wuhan and the smallest one is 5 days in Hainan \cite{wang2020clinical}. These results are incorporated in our application study.
	
	\section{Application to early outbreak of COVID-19 in China}
	
	\subsection{Trajectory datasets}
	
	\subsubsection*{Data collection and cleaning}
	We collected the trajectory data of diagnosed people and normal people in the early stage of the COVID-19 in China, including trajectories between January 1st, 2020 and January 28th, 2020 of around 9000 confirmed subscribers who were diagnosed before February 1st, 2020 and trajectories from January 17th, 2020 to January 31st, 2020 of 120 thousands normal people sampled in 25 regions of China.
	The general description of the track data field is shown in the Table \ref{data description}. We have desensitized personal data in accordance with privacy management regulations.
	
	\begin{table}[h]
		\centering
		\caption{Description of trajectory data}
		\label{data description}
		\begin{tabular}{cc}
			\toprule
			Variable name & Explanation \\
			\midrule
			User ID & Encrypted subscriber identifier \\
			District ID & Code of district where eNodeB located \\
			Lac ID & Encrypted Location Area Code identity\\
			Cell ID & Encrypted cell identity \\
			Lat & Latitude of cell \\
			Lng & Longitude of cell \\
			Timestamp & Timestamp of subscriber entering a base station cell \\
			\bottomrule
		\end{tabular}
	\end{table}
	
\begin{figure}[h]
		\centering
		\includegraphics[scale=0.5]{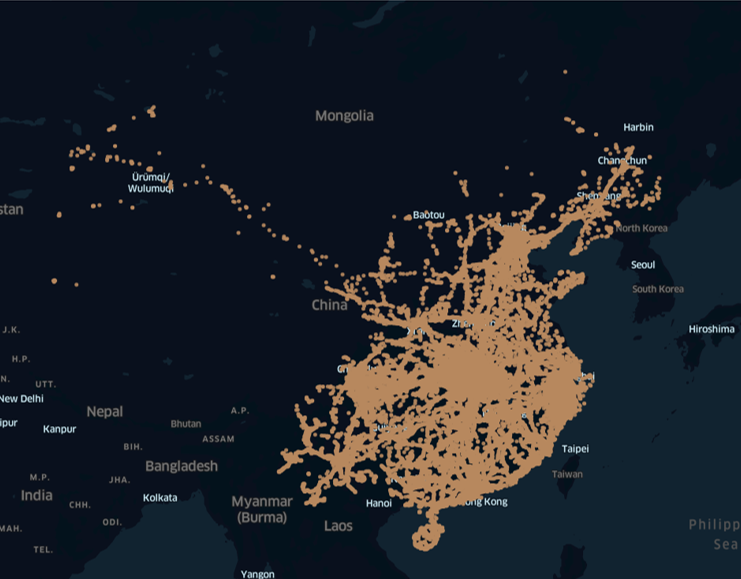}
		\caption{Distribution of the base stations cell covered by the trajectory datasets}
		\label{42stns}
	\end{figure}

	The User ID can be used as the unique identification of the subscriber, and the District ID can be used to judge the spatial range of activities. The relationship of Lac ID and Cell ID is one-to-many. The higher the population density, the smaller the base station cell signal covers \cite{ref11} and the more Cell IDs corresponding to the same Lac ID are. In this study, we use District ID\&Lac ID\&Cell ID to identify a base station cell, and the spatial distribution is shown in Figure \ref{42stns}.
	
	%
	
	Telecommunication signaling data are noisy due to its huge real-time volume. Data cleaning is necessary before we can apply our model on it. For example, wireless signals are subject to drift due to the influence of surrounding. In order to ensure seamless coverage of wireless signals and uninterrupted communication during mobile phone moving, multiple base station cells signals often overlap in one place.
	When a user’s track records switch back and forth between different base station cells in short time, which are defined as `A-B-A'-switch in short time, where A, B are base station cells, we discard the record of base station cell B of this timestamp, and believe user is still at base station cell A.
	
\begin{figure}[h]
		\centering
		\includegraphics[scale=0.5]{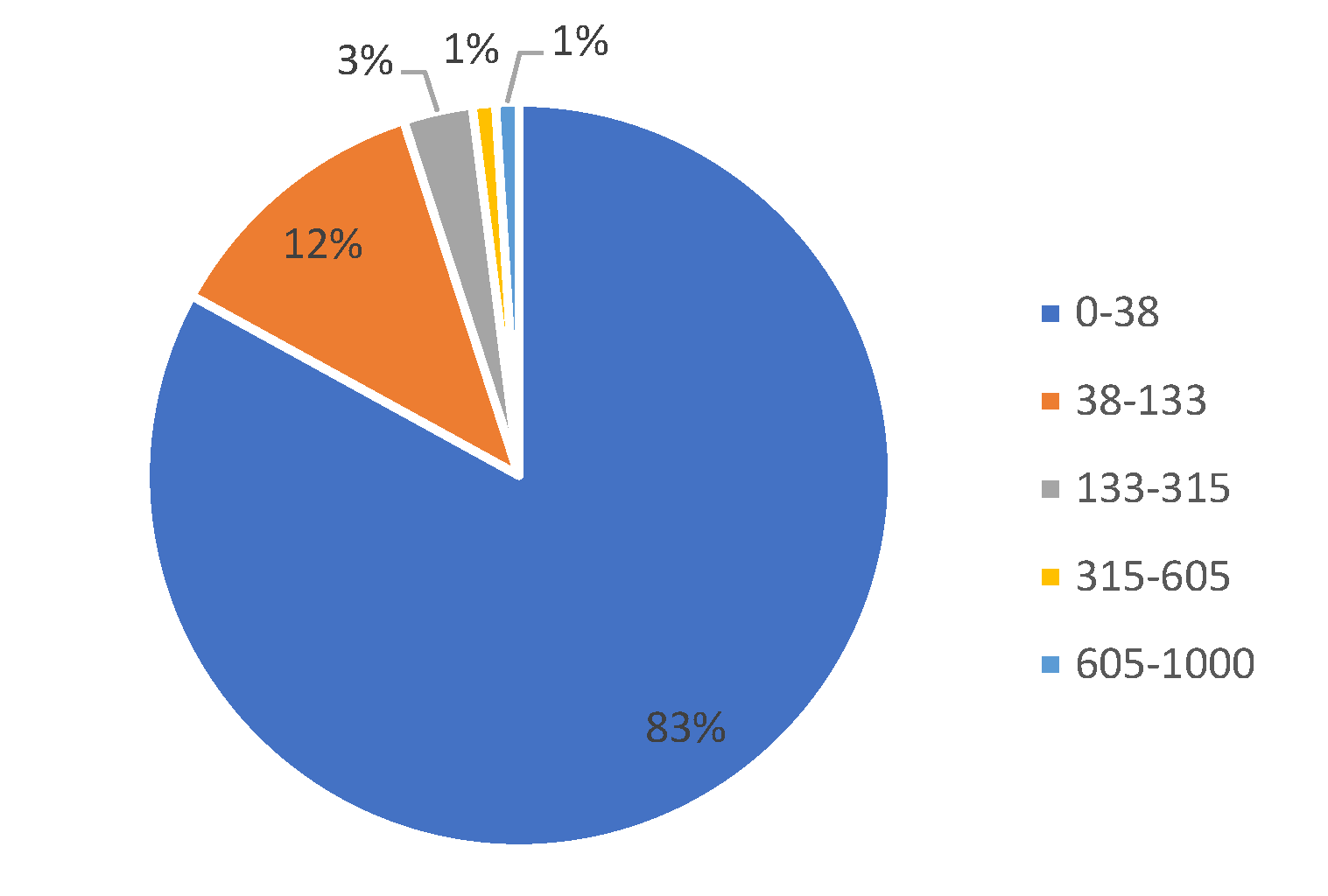}
		\caption{Clustering results of switching speed}
		\label{speed}
	\end{figure}

	The trajectory data contains information about subscriber's movement status. Using the spatial displacement and time difference of the adjacent timestamp trajectory, subscriber's historical switching speeds can be obtained. We calculate the switching speed between different base station cells and perform clustering analysis using K-means method on the effective speed. Results are given in Figure \ref{speed}.
	
	
	Clustering results using K-means method show that speeds of 83\% trajectory records are below 38km/h. We keep these records for further analysis. Exclusion of high-speed movement samples from our analysis is based on the following logic: if a confirmed case passes a BSC when traveling by car, train, airplane, etc, his viral shedding to region covered by this BSC is almost negligible by all means due to the fast movement and enclosed space. In addition, we also exclude record with dwell time less than 300 seconds assuming viral shedding needs more than 5 minutes to happen. 
	
	\subsubsection*{Data splitting}
	
	In order to evaluate our model, we divide the entire trajectory dataset of confirmed cases into two subsets according to their confirmed date. The first subset contains subscribers diagnosed before January 25th, and the second one includes subscribers diagnosed after January 25th, and the ratio of the size of these two subset is 7:2. The first dataset is used to train our model, while the second one is used to conduct cross validation and hindcast.
	
	
	\subsubsection*{Sampling of normal people}
\begin{figure}[h]
		\centering
		\includegraphics[scale=0.5]{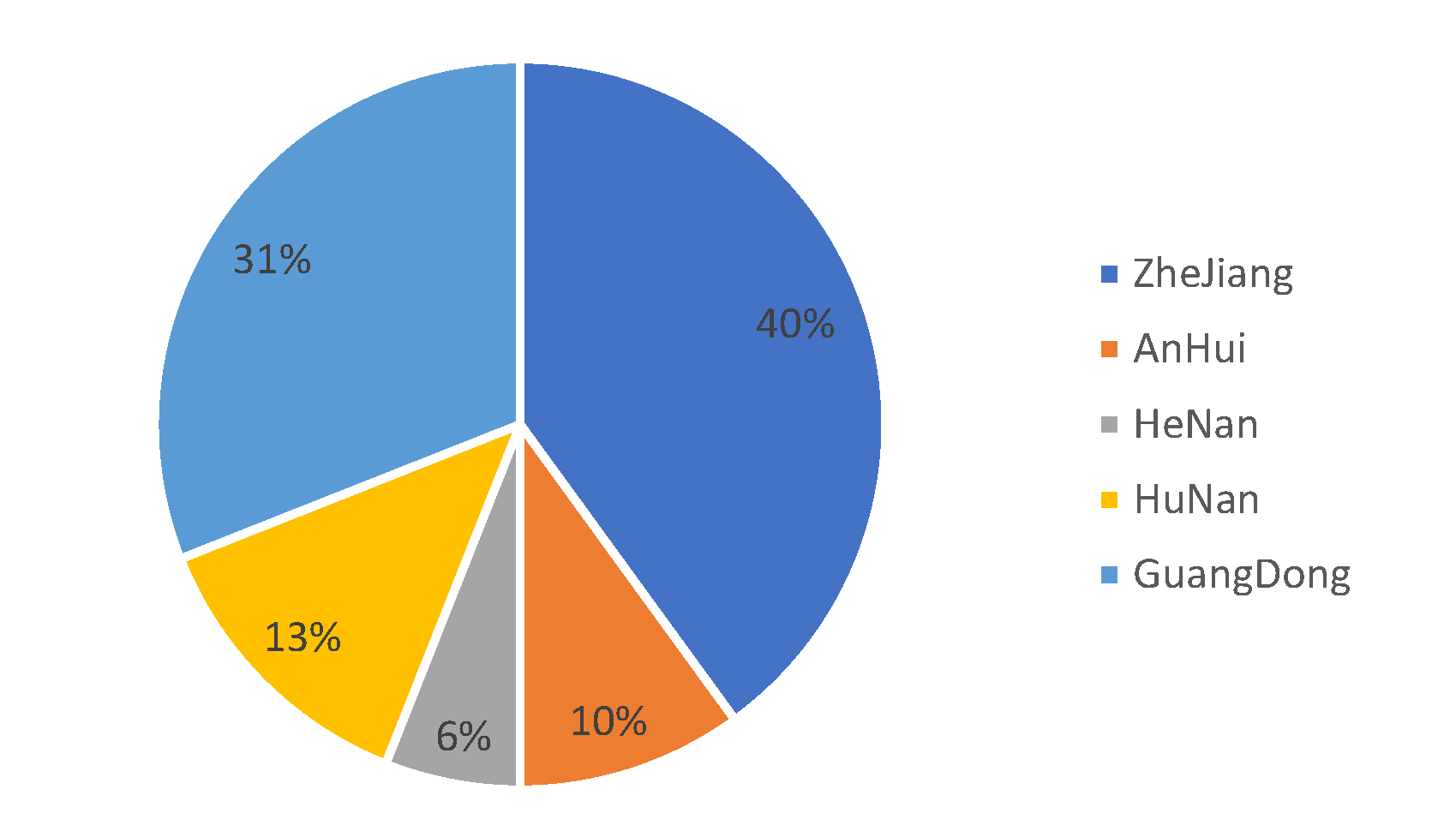}
		\caption{Spatial sampling of normal subscribers}
		\label{sampling ratio 2}
	\end{figure}

	Due to the huge amount of normal samples and the distinctive difference between number of confirmed and health cases, we use stratified sampling to randomly sample a total of 120 thousands subscribers in 25 regions, and results of sampling ratio of normal subscribers are shown as Figure \ref{sampling ratio 2}. And we employ the same trajectory data extraction method of diagnosed people and obtain proportion of dwell time of base station cells to evaluate our model.
	
	\subsection{Assessment of COVID-19 risk maps in China}
	
\begin{figure*}
		\centering
		\subfigure[City-level risk]{\includegraphics[height=1.5in]{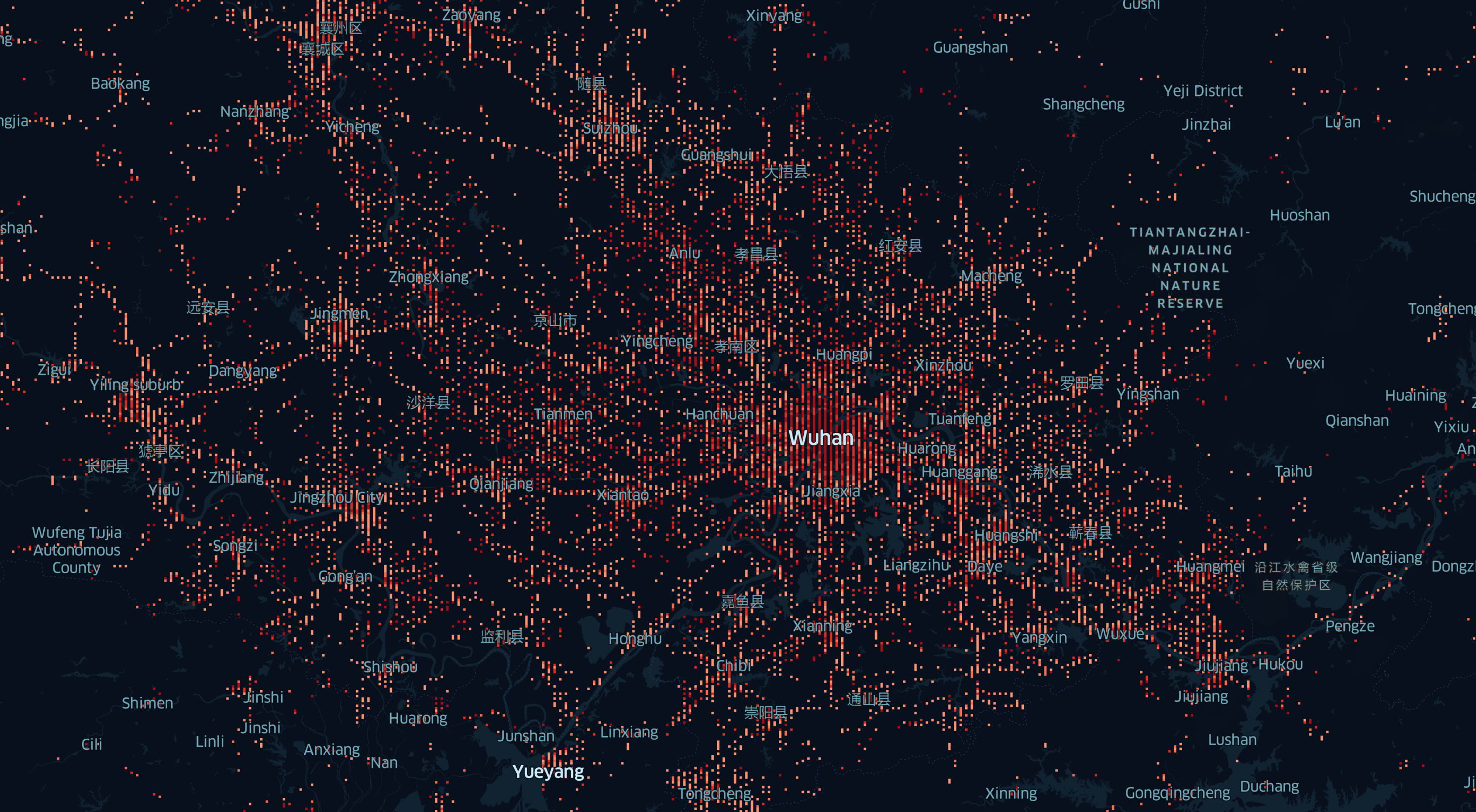}}
		\subfigure[Community-level risk]{\includegraphics[height=1.5in]{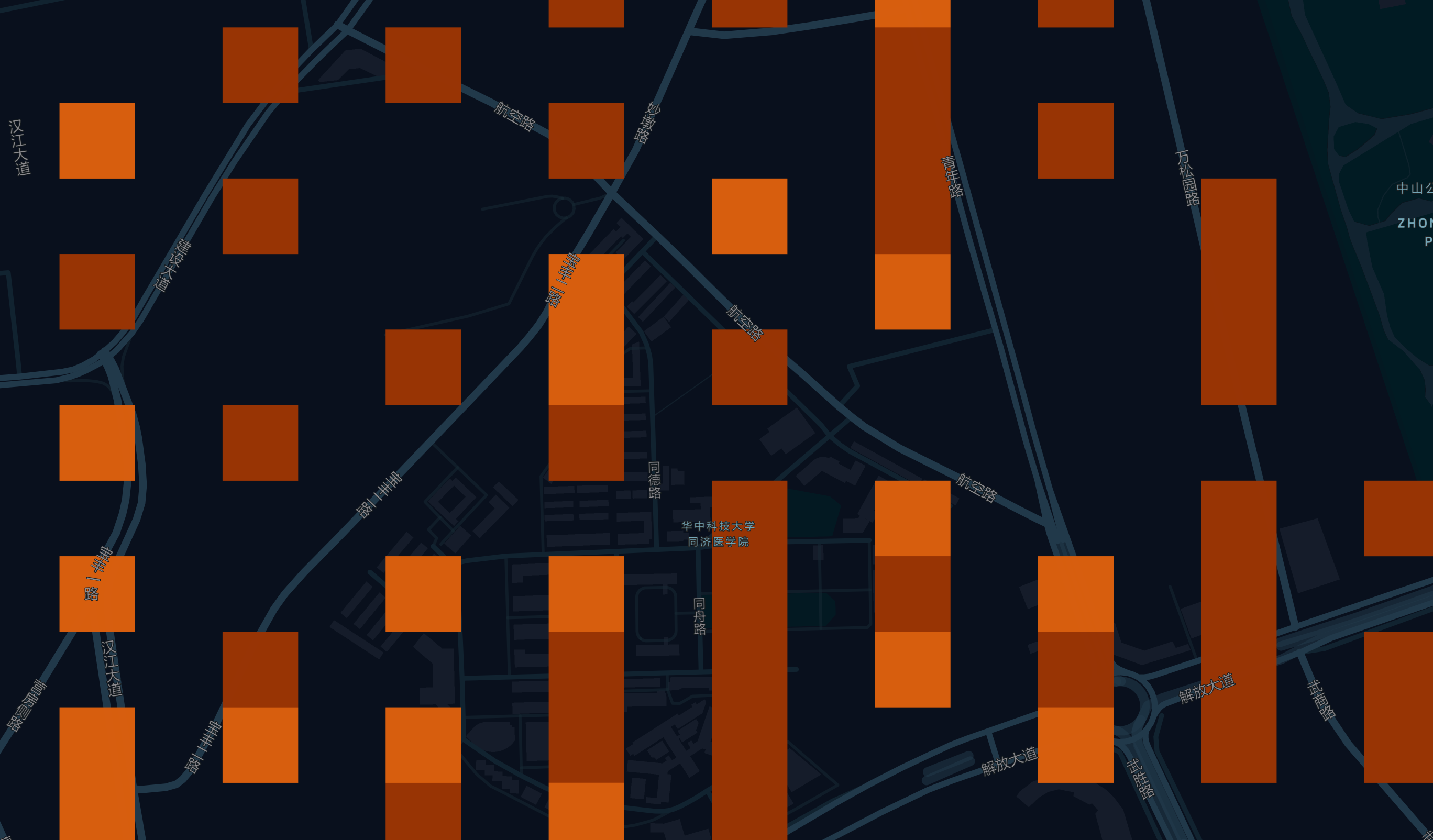}}
		\caption{Visualization example of HiRES risk map}
		\label{riskmap0127}
	\end{figure*}

	We calculate COVID-19 risk map in China using HiRES model and the trajectory dataset described in the previous subsection for the period of January 1st to January 27th, 2020. These maps cover over 100,000 communication base station cells each day, and the radius of each BSC range from 500 meters to 1000 meters, approximately. Thus our HiRES risk map has a dynamic spatial resolution, ranging from 500 meters to 1000 meters. The risk map of January 27th is shown in Figure \ref{riskmap0127} as an illustration. It can be observed that Wuhan urban area has the highest risk value, and risk of infection radiates to surrounding cities and counties with Wuhan as the center. This is consistent with the situation that, in China, COVID-19 firstly outbreak in Wuhan in January and spread rapidly to the surrounding cities and counties, indicating that HiRES risk map is able to capture the macro spatial pattern of COVID-19's spreading risk.
	
	In addition, we analyze the correlation between the daily total confirmed cases and the regional averaged risk in the HiRES risk map from January 20th to 28th over China, Beijing, and Wuhan, respectively. The estimated correlation coefficients for China, Beijing, and Wuhan are $0.9492$, $0.9665$ and $0.9967$. This reveals that the regional risk estimated by the HiRES risk map has a strong positive correlation with the total number of confirmed cases in that region. As is shown in Figure 6-(a), from a macro perspective, the HiRES risk map can reasonably estimate the total regional risk and the size of the infected population, thus can be used as a scientific support for the macro policy-making of the epidemic prevention and control.
	
	We also investigate local temporal variability of representative locations in Wuhan and Beijing. Six categories of locations are investigated, which are hospital, commercial district, station, school, residence and randomly selected area. 10 samples are collected for each category in each city. Totally 120 locations are evaluated. Based on the daily HiRES risk map, the epidemic risk of each location is calculated as the sum of the risk values of the base station cells within the specified radius of the location, and radius of each location is set according to actual conditions for each category as specified in Table \ref{loc_distance}.
	
	\begin{table}[h]
		\centering
		\caption{Coverage radius of location category}
		\label{loc_distance}
		\begin{tabular}{ll}
			\hline
			Location category & Coverage radius  \\
			\hline
			Hospital &  300m\\
			Market &  100m\\
			School &  20-500m \\
			Station &  500m\\
			Residence & 100m\\
			Random area & 100m\\
			\hline
		\end{tabular}
	\end{table}
	
	
	\begin{figure*}
		\centering
		\subfigure[Regional evolution of epidemic risk in China]{\includegraphics[height=1.5in,width=3in]{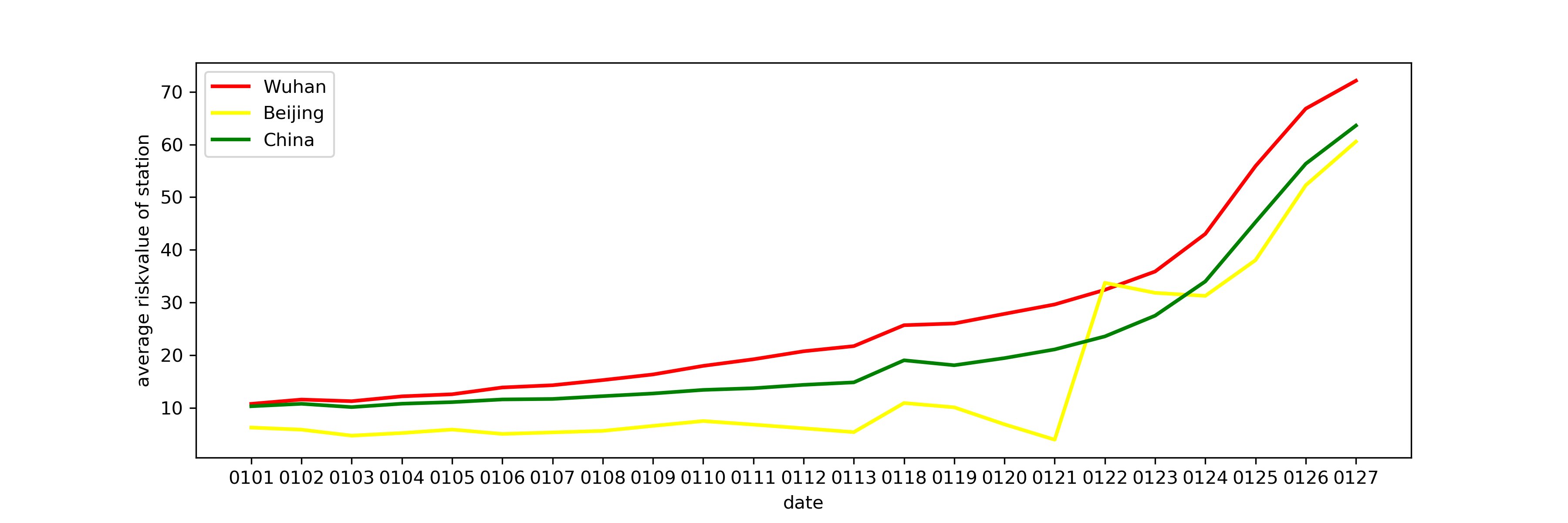}}
		\subfigure[Local evolution of epidemic risk in Beijing]{\includegraphics[height=1.5in,width=3in]{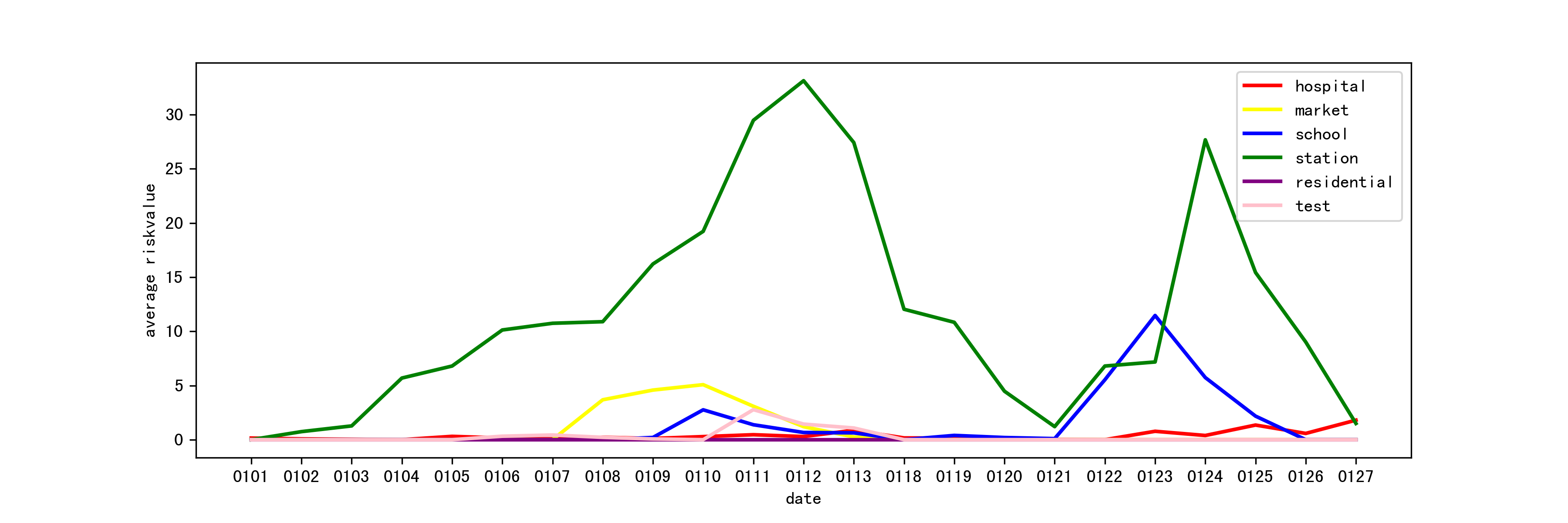}}

		\subfigure[Local evolution of epidemic risk in Wuhan]{\includegraphics[height=1.5in,width=3in]{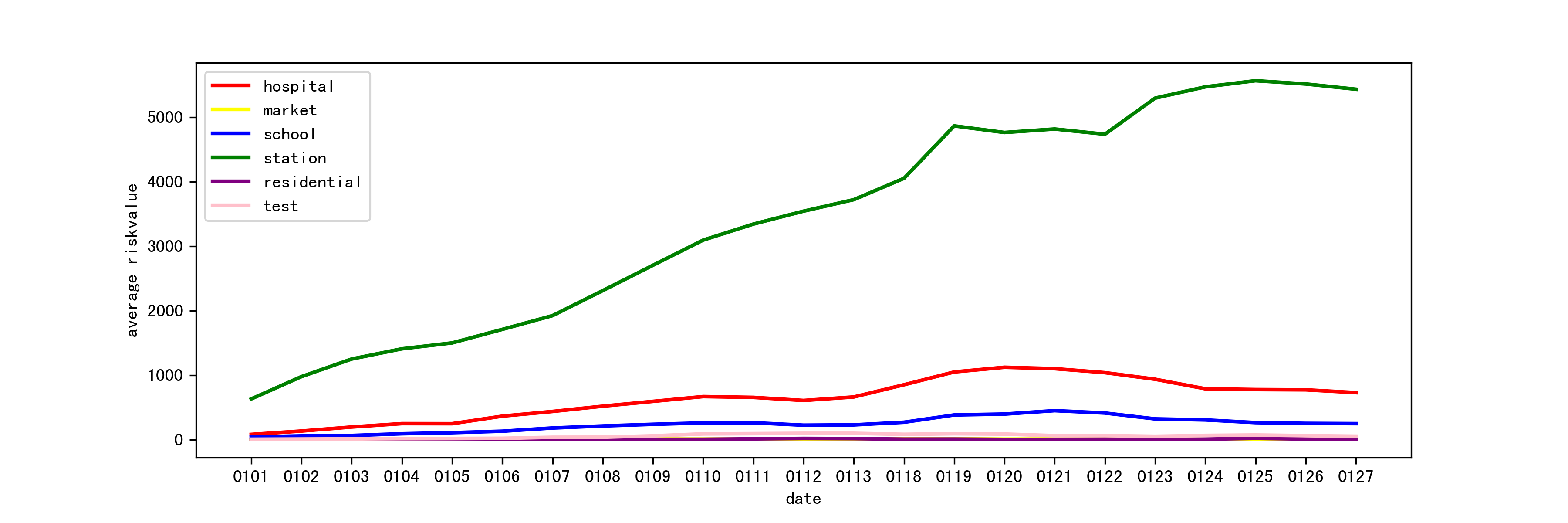}}
		\subfigure[Evolution of epidemic risk in local markets]{\includegraphics[height=1.5in,width=3in]{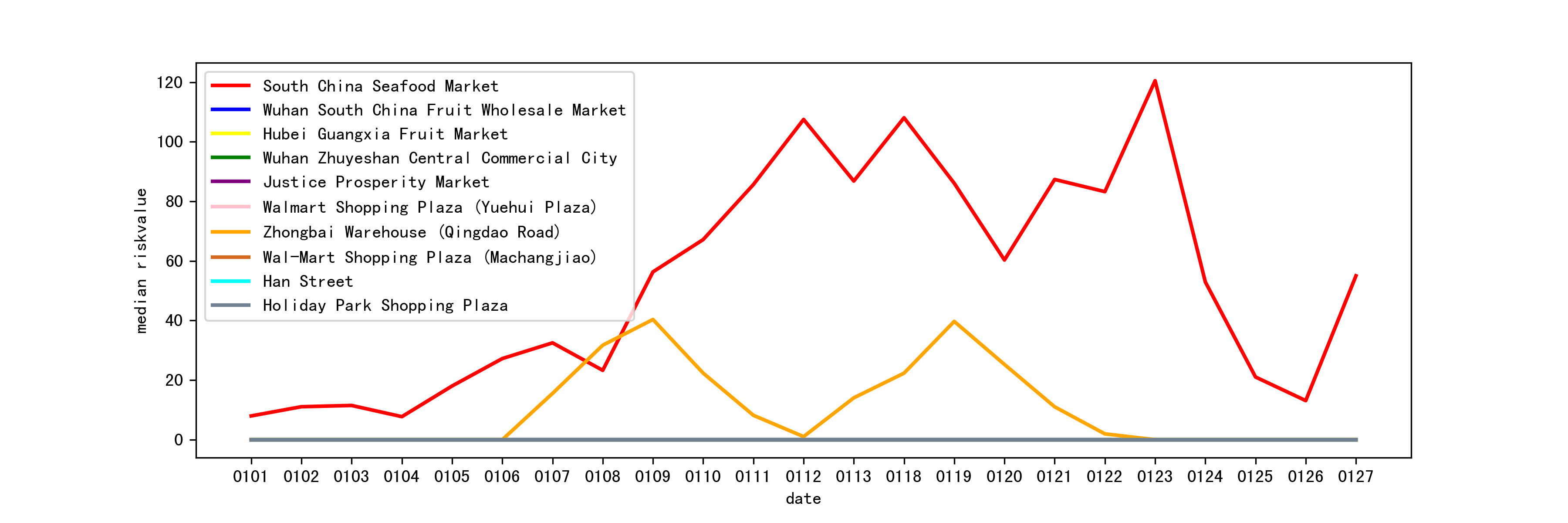}}
		\caption{2020/1/01-2020/1/27 Evolution of epidemic risk}
	\end{figure*}
	
	Figure 6-(b) and (c) shows the temporal evolution of COVID-19 risk estimated by the HiRES map for each category of local locations in Beijing and Wuhan, respectively. It's clearly observed that in January 2020, local risk evolution in Beijng and Wuhan are totally different, with those in Wuhan exhibiting a significant trend of increase. But location category of high-risk identified from the HiRES risk map are consistent between Beijing and Wuhan, which are station, hospital and school. These findings coincident with common sense and the actual situations since these are all locations with dense crowds and infected cases often gather. 
	
	In addition, among the ten designated hospitals for COVID-19, Tongji Hospital of Huazhong University of Science and Technology, Wuhan Union Hospital, and Wuhan Pulmonary Hospital have higher risk values. Among them, Risk of Tongji Hospital is several times higher than other hospitals, which is confirmed by the fact that Tongji admitted the most confirmed cases during the early outbreak in Wuhan. In the category of supermarket, the Southern China Seafood Market in Wuhan, as a well known gathering place of confirmed cases in the early stages of the outbreak, has been singled out with consistently and significantly higher risk estimate as shown in Figure 6-(d); its risk started to increase rapidly from January 4th and reached a peak around mid of January, then experienced a sharp decline after the closure of the city.
	
	During its early outbreak, the COVID-19 risk in Beijing is mostly identified in stations as suggested by the HiRES maps. No trend of increase observed in January indicates that transmission of the virus were sporadic at that time. Two risk peaks are identified in Beijing's stations in January 12th and January 24th, which probably corresponds to the first risk peak observed in Wuhan Southern China Seafood Market and risk caused by the traffic flow out of the Wuhan city by locking down, respectively.

\begin{figure*}
		\centering
		\subfigure[Boxplot of 10 stations at Beijing]{\includegraphics[width=3in]{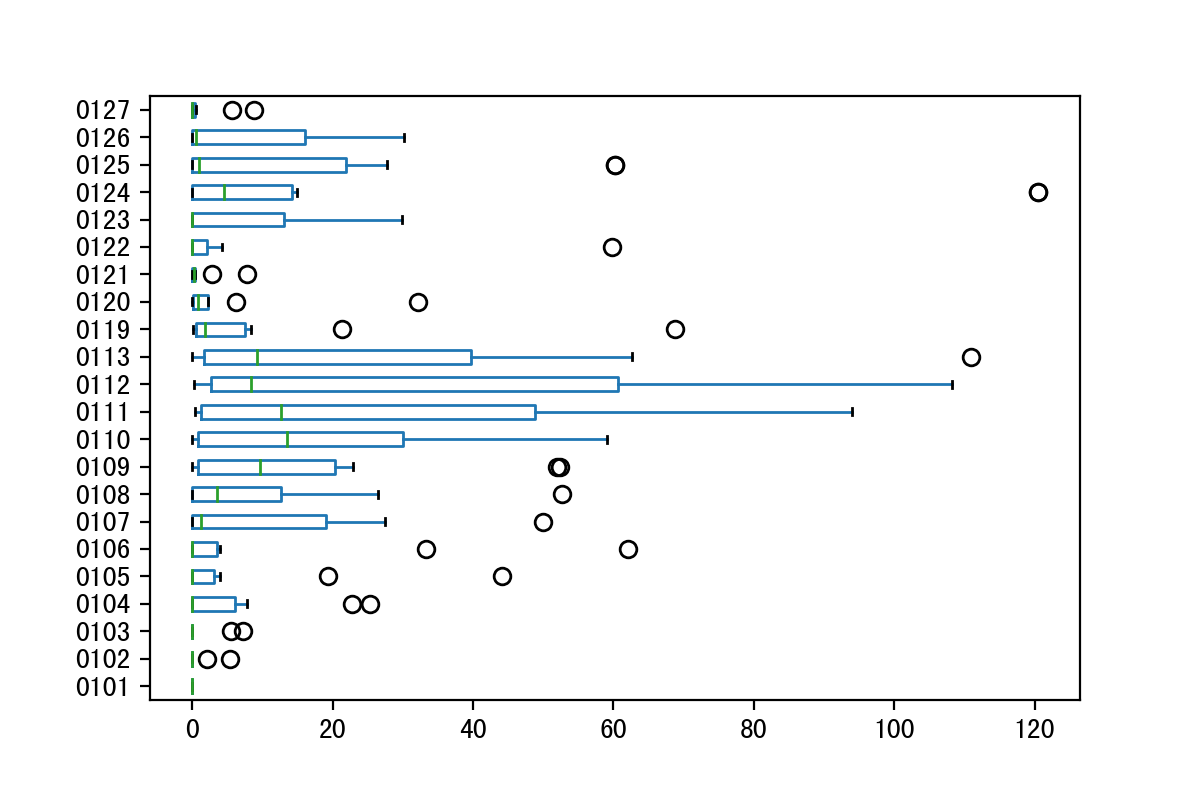}}
		\subfigure[Boxplot of 10 hospitals at Wuhan]{\includegraphics[width=3in]{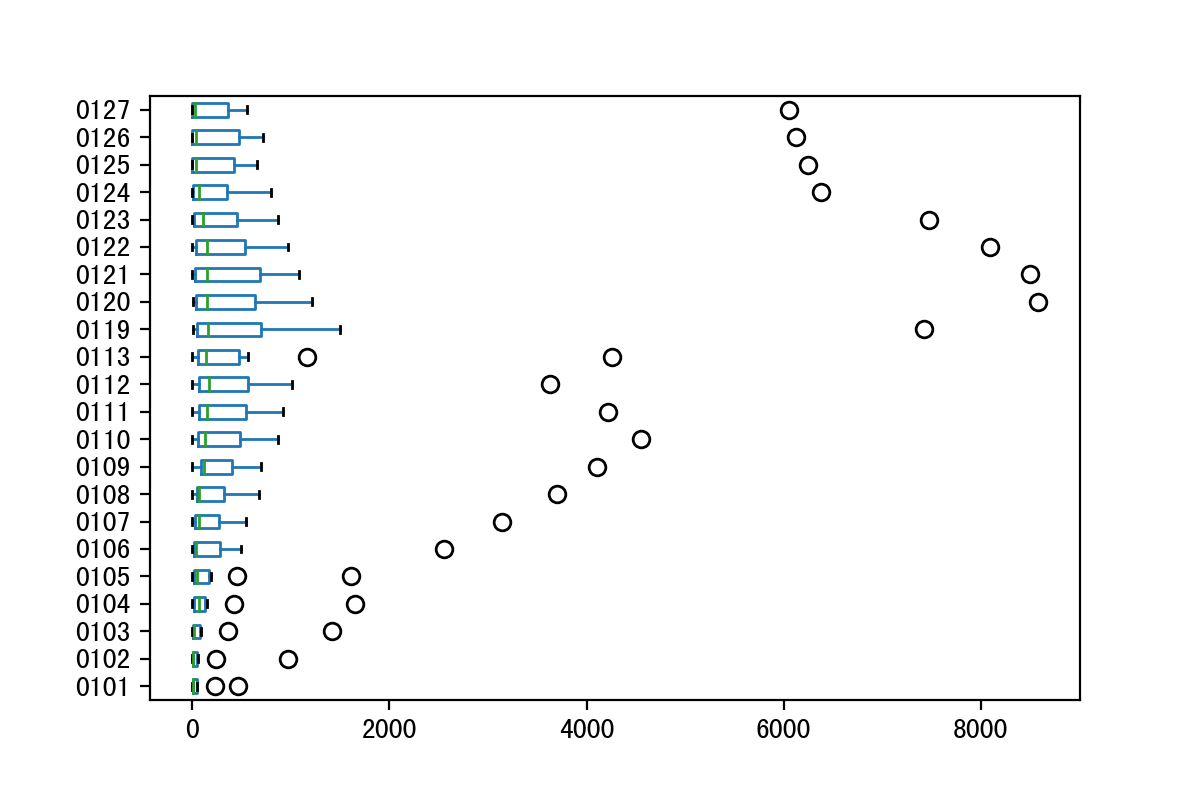}}
		\caption{ Spatio-temporal variability of  COVID-19 epidemic risk given by HIRES risk map }
		\label{boxplot}
	\end{figure*}

	We further investigate the spatio-temporal variability of the epidemic risk of COVID-19 in China based on the HiRES risk maps. We use the box-plot of daily HiRES risk estimates to visualize the daily spatial variability of COVID-19 epidemic risk during January 1st to January 27th, 2020 within each location category. Results show that spatial variability of the COVID-19 epidemic risk usually increase as the risk values increase. In other words, when the risk of COVID-19 epidemic increased, the risk changes within each category of high-risk locations such as hospital and station became larger as shown in Figure \ref{boxplot}. Thus real-time high-resolution risk monitoring is indispensable in order to achieve efficient deployment of medical and social resources, especially when the epidemic risk is spreading fast. 
	
	Above all, the HiRES risk map is shown to be capable of providing accurate estimations of the COVID-19 epidemic risk in China during its early outbreak, at both macro and micro scales. The spatio-temporal variability of the COVID-19 epidemic risk is large based on our HiRES risk maps. Therefore, high-resolution quantification of epidemic risk is necessary to guide the policy-making and practice of epidemic prevention and control in order to achieve efficient and economic prevention and control effects.

	\subsection{Detection of suspected individuals}
	
	Based on the HiRES risk maps, we calculate the HiRES-p score for the 2757 confirmed cases in the training dataset and for 100000 normal subscribers obtained from the sampling dataset for comparison. We apply both statistical methods and machine learning methods to detect suspected cases in this mixed dataset, which represents scenario with 3\% population infection rate.
	
	To conduct hypothesis testing, we firstly derive the cumulative distribution function under null hypothesis. $F_0$ is estimated by the empirical cumulative distribution function $\hat{F}_0$ of 100000 HiRES-p scores calculated for the normal group on January 24th, 2020. Consequently the $95\%$ quartile of $\hat{F}_0$ is taken to be the critical value of the test, which is 1.19955.  Thus the significant level of the test is expected to be 5\%, or in other words, we control the type I error of false alarm at 5\%. Then, we calculate testing statistic for each individual in the training dataset, i.e., their HiRES-p scores, and compare them with the critical value. If the HiRES-p score is larger than the critical value, this individual is declared to be detected as a suspected case. A total of 2186 are detected out of 2757 confirmed cases, indicating an overall detection rate around 80\%. Detection rates with respect to the date of diagnosed are presented in the following Figure \ref{5day-Htest}. When controlling the false alarm rate to be $5 \%$, the detection rate of our statistical inference based method is about $80 \%$ with low temporal variability during the 5-day period.
	
	\begin{figure}[h]
		\centering
		\includegraphics[scale=0.7]{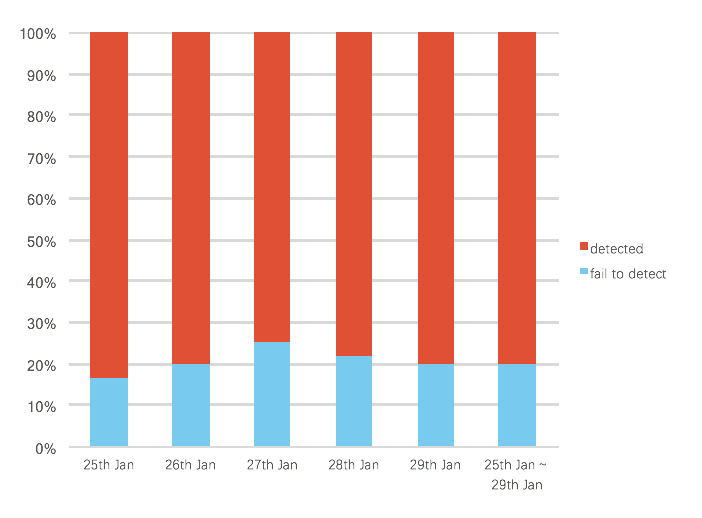}
		\caption{Detection rate of statistical inference method}
		\label{5day-Htest}
	\end{figure}

	For machine learning based detecting method, we firstly extract data feature by defining an explicit function form for calculating HiRES-p score.  Since machine learning method is good at dealing with high dimensional data, it's not necessary to reduce the dimension of the score series during the incubation period such as through summation or taking maximum. If we assume the incubation period parameter $T=14$, a simple feature would be a $T$-dimension vector containing the $T$ HiRES-p scores calculated during the past incubation period. Unfortunately, trajectory data for the normal group only start from January 17th. Thus, due to limitation of our dataset, our machine learning algorithms use only 8 features which are the HiRES-p scores obtained from January 17th to January 24th. This is a classical binary classification problem, and we assume ``diagnosed" to be 1 and ``healthy" to be -1. We select $80\%$ of the mixed dataset as the training set with the rest of them to be the testing set. We try three classical classifiers: Support Vector Machine, Decision Tree and Random Forest. A successful detection is defined as a correct classification. We evaluate the performance of the machine learning based method in terms of accuracy (ACC) and results are given in Table 3.

	\begin{table}[]
		\centering
		\caption{Accuracy of machine learning detection method}
		\begin{tabular}{llll}
			\hline
			Population Infection rate & Algorithm               & ACC & Computing time(s) \\
			\hline
			\multirow{3}{*}{1\%}      & Support Vector Machine & 99.11\%                            & -                 \\
			& Decision Tree           & 98.86\%                            & 0.9199            \\
			& Random Forest           & 99.11\%                            & 1.65              \\
			\hline
			\multirow{3}{*}{2\%}      & Support Vector Machine & 98.48\%                            & 618.91            \\
			& Decision Tree           & 97.92\%                            & 0.4199            \\
			& Random Forest           & 98.47\%                            & 0.77              \\
			\hline
			\multirow{3}{*}{3\%}      & Support Vector Machine & 98.01\%                            & 196.57            \\
			& Decision Tree           & 97.43\%                            & 0.2900            \\
			& Random Forest           & 98.03\%                            & 0.56              \\
			\hline
			\multirow{3}{*}{10\%}     & Support Vector Machine & 94.74\%                            & 37.84             \\
			& Decision Tree           & 93.75\%                            & 0.089             \\
			& Random Forest           & 95.06\%                            & 0.17              \\
			\hline
			\multirow{3}{*}{15\%}     & Support Vector Machine & 92.34\%                            & 22.66             \\
			& Decision Tree           & 91.45\%                            & 0.060             \\
			& Random Forest           & 93.07\%                            & 0.1099            \\
			\hline
			\multirow{3}{*}{23\%}     & Support Vector Machine & 88.52\%                            & 10.0400           \\
			& Decision Tree           & 89.14\%                            & 0.0300            \\
			& Random Forest           & 91.09\%                            & 0.0799            \\
			\hline
			\multirow{3}{*}{50\%}     & Support Vector Machine & 74.79\%                            & 4.2400            \\
			& Decision Tree           & 84.96\%                            & 0.0300            \\
			& Random Forest           & 84.12\%                            & 0.0600           \\
			\hline
		\end{tabular}
		\label{dr-ml}
	\end{table}
	
	For statistical inference based method, detection rate is defined by successful identification of suspected cases. For comparison purpose, we transform the detection rate of the inference based method to a classification accuracy estimation, i.e., $(2186+95000)/(2757+100000) \approx 94.58 \%$. Therefore, we conclude machine learning based method slightly outperforms the inference based method in terms of detection accuracy when population infection rate is $3\%$. 
	
\begin{figure}[h]
		\centering
		\includegraphics[scale=0.5]{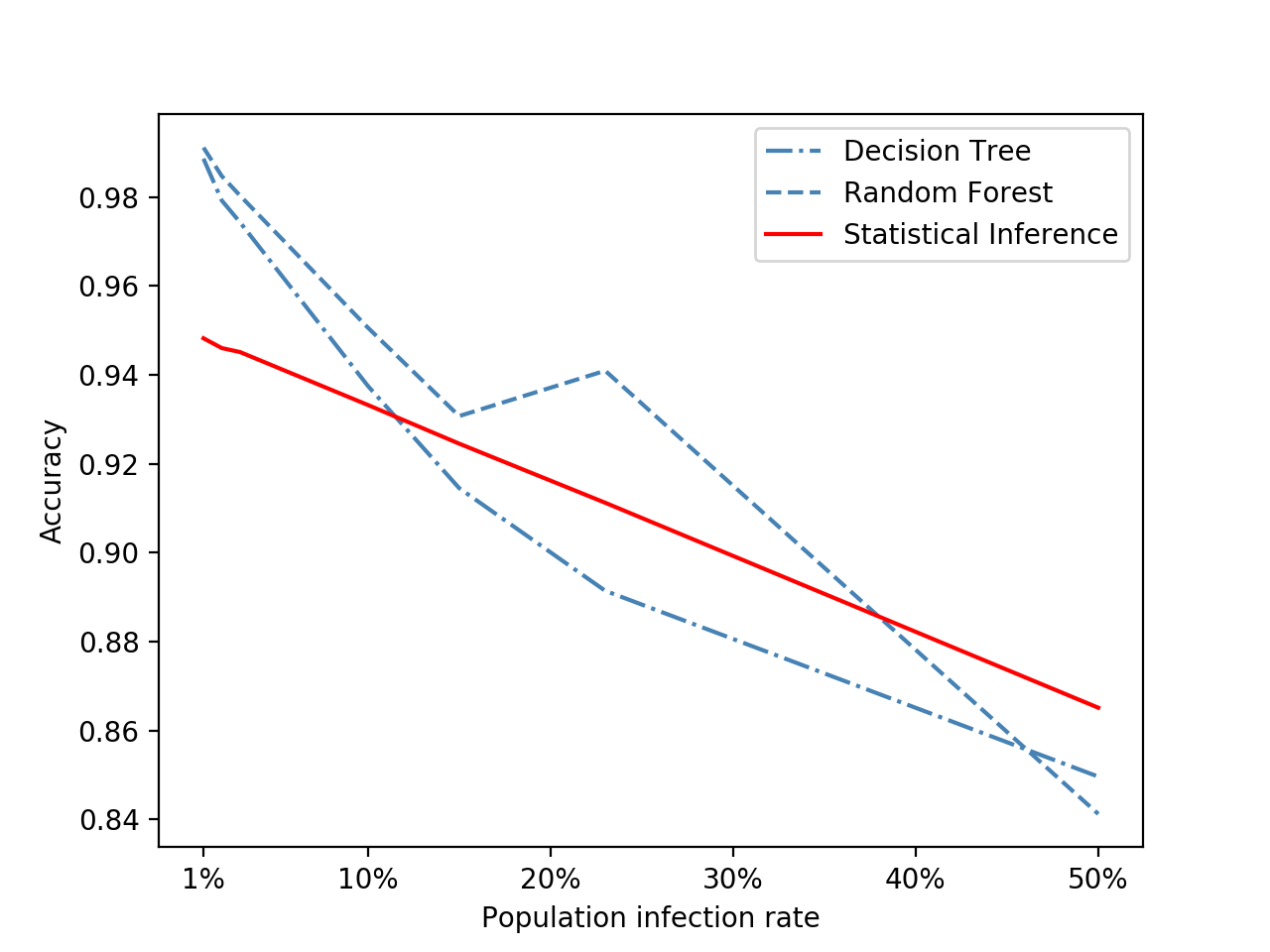}
		\caption{Accuracy of detection v.s. Rate of population infection}
		\label{cr-comp}
	\end{figure}

	Observed situations in China and recent research have reported different population infection rate of COVID-19, ranging from 1\% to 30\% \cite{zheng2010understanding}. We design more numerical experiments to simulate scenarios with different population infection rate ranging from 1\% to 50\% by changing the size of the normal group. The accuracy of classification is shown in Figure \ref{cr-comp}. Negative linear relationship is observed between our accuracy and the rate of population infection. The performance of machine learning based method is better than statistical inference based method when the population infection rate is below 10\%, true in most cases; while the inference based method outperforms machine learning based method as rate of population infection increases. In general, the statistical inference based method is more robust.

	In summary, based on our numerical experiments, HiRES-p score has been proven to be an accurate measurement of individual's epidemic risk for COVID-19, and can be used to detect suspected cases. Machine learning based methods in general outperform the statistical inference based method in terms of detection accuracy under scenarios close to the real world; while the statistical inference based method is more robust. If the population infection rate is controlled below 20\%, accuracy of classification is always above 90\%.
	
	We use Python3.7 to conduct numerical experiments on server with 4 cpus, cpu Core: 8, model: Intel(R) Xeon(R) CPU E5630 @ 2.53GHz. It takes about 0.75 seconds per 10000 pieces of trajectory data collected to get the HiRES risk map, and about 128 seconds per 10 thousand individuals to obtain the HiRES-p scores. Therefore if sufficient computing power is provided, reporting-lag of the HiRES risk map can be easily controlled to an hour and minute-level update of HiRES-p score is also feasible. 
	
	\section{Conclusion and discussion}
	\label{sec5}
	Based on the trajectory data of diagnosed cases and the mean field theory, we propose a high-resolution spatio-temporal model for epidemic risk assessment, with a fine and dynamic spatial resolution. Using a series of epidemic risk maps produced by HiRES model, we develop the objective personal risk scoring model, i.e., HiRES-p, to obtain explicitly quantified risk of infection for every authorized individual. Based on HiRES-p scores, we adopt statistical inference approach and machine learning approach respectively, to predict early infection of suspected individual. Models and methods are applied to investigate the epidemic risk in China during the early outbreak of 2019 Novel Coronavirus, i.e., January 1st, 2020 to January 28th, 2020.
	
	Results show that HiRES risk maps can capture the spatio-temporal pattern of COVID-19 epidemic risk at both macro and micro scale with satisfactory accuracy. The regional risk estimated by the HiRES maps are highly positively correlated to the total number of confirmed cases in that area. High-risk locations such as hospitals, stations and the Southern China Seafood Market in Wuhan can be easily distinguished by using their risk values on the HiRES map. The spatio-temporal variability of the COVID-19 epidemic risk is shown to be large based on our HiRES risk maps, and the variability increase as risk values increase. Therefore, high-resolution quantification of epidemic risk is necessary to guide the policy-making and practice of epidemic prevention and control in order to achieve efficient and economical prevention and control effects as well as efficient deployment of medical resources. 
	
	We conduct numerical experiments using trajectory data of COVID-19 confirmed cases and normal individuals to exam the detecting ability of our HiRES-p model for suspected cases. We employ both statistical inference based method and machine learning based method. In our experiment, the accuracy is linearly and negatively related to the population infection rate. The performance of both detecting methods in terms of accuracy are in general above 90\% under scenarios with different population infection rate ranging from 1\% to 20\%.  The performance of machine learning based method is better than inference based method when the population infection rate is below 10\%, true in most cases; while the inference based method outperforms machine learning based method as rate of population infection increase. In general, the inference based method is more robust. Results indicate great application potential in epidemic risk prevention and control of HiRES-p methodology. The trajectory data we used is from one of the three major telecommunication carriers in China. It can only provide the trajectory data for a sample of confirmed case population, which results in lack of coverage of the HiRES risk map. Thus the detection capability of HiRES-p may be further improved if the trajectory data for all confirmed cases is available.
	
	The COVID-19 epidemic most possibly will not vanish in short time, and may present periodic pattern of outbreaks if the vaccine cannot be successfully developed. We need more objective data-driven tools such as HiRES and HiRES-p to monitor its evolution over space and among population. In the future, trajectory big data can be employed to automatically identify the route of confirmed and suspected cases, to verify the consistency of the patient claims collected through epidemiological investigation, as well as to conduct traceability analysis of epidemic disease. In addition, through joint application of the trajectory data and personal health data,  more accurate detection method could be developed since we simultaneously take internal and external factors, i.e., the individual's physical health condition and external exposure situations, into account.

	\paragraph*{Acknowledgements} This study was supported by the National Natural Science Foundation of China (Grant No.11421101).

	\bibliographystyle{unsrt} 
	\bibliography{HiRES}


\end{document}